\documentclass[pra,twocolumn,showpacs,superscriptaddress,floatfix]{revtex4-1}
\usepackage{epsfig}
\usepackage{latexsym}
\usepackage{amsmath}
\usepackage{graphics}
\usepackage{graphpap}
\usepackage{bm}
\usepackage{graphicx}

\begin{document}
\title{Quantum catastrophes and ergodicity in the dynamics of bosonic Josephson junctions}
\author{D. H. J. O'Dell}
\affiliation{Department of Physics and Astronomy,
    McMaster University, 1280 Main Street W.,
     Hamilton, Ontario, Canada, L8S 4M1}

\begin{abstract}
We study  rainbow (fold) and cusp catastrophes that form in Fock space following a quench in a Bose Josephson junction. In the Gross-Pitaevskii mean-field theory the rainbows are singular caustics, but in the second-quantized theory a Poisson resummation of the wave function shows that they are described by well behaved Airy functions.  The structural stability of these Fock space caustics against variations in the initial conditions and Hamiltonian evolution is  guaranteed by catastrophe theory. We also show that the long-time dynamics are ergodic. Our results are relevant to the question posed by Berry [M.V. Berry, Nonlinearity \textbf{21}, T19 (2008)]: are there circumstances when it is necessary to second-quantize wave theory in order to avoid singularities?  
\end{abstract}

 \maketitle

In the geometrical theory of light, rainbows are caustics formed by the focusing of the Sun's rays by raindrops. Caustics are the envelopes of families of rays and the light intensity diverges upon them: they are catastrophes where the ray theory fails. In order to tame the divergences it is necessary to go to the next level of sophistication, namely the wave theory of light. For rainbows this problem was solved by Airy in 1838 \cite{airy}.

In this Letter we investigate caustics that form in \emph{Fock space}  during quantum many-body (QMB) dynamics. The analog of ray and wave theory are   mean-field (MF) and QMB theory, respectively. Caustics in Fock space  are places where a classical field must be second quantized in order to remove  divergences.   Just as wave theory introduces the notion of phase, QMB theory introduces second quantization, i.e.\ discreteness of particle number, and this smooths the singularity.  There are many situations where MF theory is quantitatively inaccurate,  but at its caustics it fails qualitatively, predicting infinities, and second quantization is forced upon us. 

Related ideas have previously been presented in the context of phase singularities in light waves \cite{leonhardt02,berry04}. By contrast, the Fock space caustics discussed here are amplitude singularities.  Caustics also occur in atom optics,  e.g.\ during the diffraction of atoms by a standing wave of light \cite{odell01} as observed in the experiment  \cite{huckans}, and during the expansion of a Bose-Einstein condensate (BEC) \cite{chalker09}. However, these latter studies concerned caustics in real or momentum space where matter wave interference from first quantization smooths the caustic singularity. 

Cold atoms are useful for studying quantum dynamics because of their tunability and long coherence times \cite{polkovnikov10}. A key experimental workhorse for such studies is the degenerate Bose gas in an optical lattice \cite{greiner2002,greiner2002b,tuchman06,strabley2007,will2010,trotzky2011,chen11}.  The system we consider here is a Bose Josephson junction (BJJ) made of two lattice sites. Two pioneering BJJ experiments \cite{albiez05,levy07} have observed coherent macroscopic phenomena, including the ac and dc Josephson effects and  self-trapping \cite{milburn97,smerzi97}. The effect of changing the height of the tunneling barrier upon Josephson dynamics has also been measured \cite{leblanc11}. These experimental advances have stimulated the development of sophisticated theoretical methods \cite{sakmann09,martinez09}. 

\begin{figure}[t]
{\includegraphics[width=1.0\columnwidth]{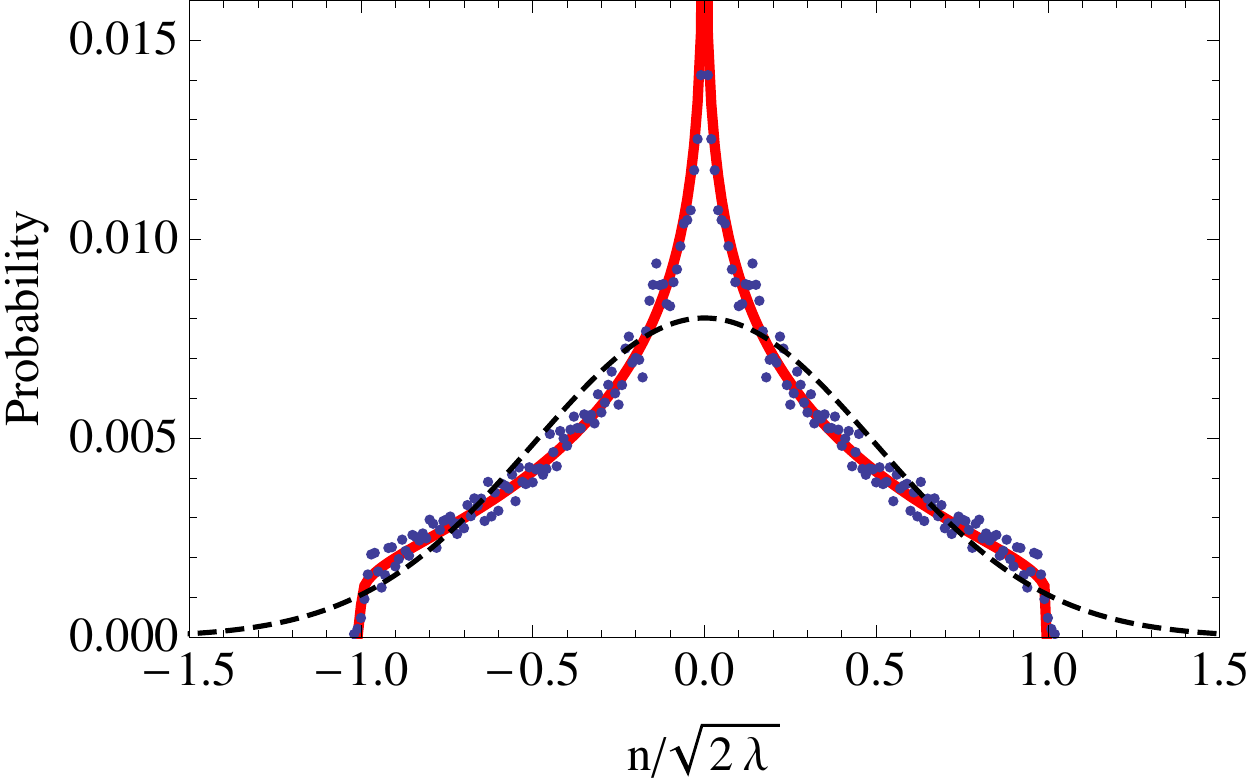}}
\caption{Ergodicity at long times. Solid red curve: microcanonical result for the number difference probability distribution [Eq.\ (\ref{eq:ergodicprobability}).] Blue dots: time average of the QMB distribution over range $\omega_{\mathrm{pl}}t=21.2\pi$--$22.2 \pi$ with $\lambda=5000$.  Dashed black curve: QMB distribution at thermal equilibrium.}\label{fig:ergodic}
\end{figure}

Here we calculate the dynamics of a BJJ following a sudden quench (increase) in the tunneling amplitude. This situation is related to the classic problem of the connection of two separate BECs \cite{zapata03,xiong06,martinez09}. The quench results in a periodic collapse and revival of the atom number distribution between the two sites \cite{milburn97,tuchman06,chuchem10}. We identify each revival with  the birth of a pair of rainbows in Fock space: the rainbows proliferate with time, giving rise to a characteristic shape for the long-time number distribution (see Fig.\ \ref{fig:ergodic}). In the macroscopic regime of large atom number, the rainbows dominate the QMB wave function and are singular in the MF theory. The proper mathematical description of rainbows is via catastrophe theory \cite{thom,berry81}: it prescribes the shape of the wave function at a rainbow and also guarantees that rainbows are \emph{structurally stable} i.e.\ stable to perturbations in the initial conditions or in the Hamiltonian governing the time evolution. Below we use catastrophe theory to understand singularities in the many-body wave function.  
%catastrophe theory and is therefore both generic and structurally stable , i.e.\ qualitatively unchanged by variations in the initial conditions or the Hamiltonian which governs the time evolution. In fact, the local structure is determined only by the 1+1 dimensions of Fock space plus time, which is the control space in which the wave function lives. 

%collapses and revivals in the distribution of atom number differences between the two sites that follow a sudden increase in the tunneling amplitude for an initially balanced number distribution. We show that each revival corresponds to the birth of a QMB rainbow. This is related to the classic problem of the connection of two separate BECs \cite{zapata03,xiong06,martinez09}, and, indeed, we shall evolve from the same initial state (a single relative Fock state), although due to structural stability (see below), this is not a requirement for the generation of rainbows. 

%Reference \cite{milburn97}  predicted a collapse and revival sequence in the distribution of atom number differences between the two sites if all the atoms are initially localized on one site. 

In the tight-binding regime the BJJ obeys the Bose-Hubbard (BH) Hamiltonian \cite{leggett01,P+S,gati07}
\begin{equation}
%H=\frac{1}{2}E_{c}n^2-E_{J} \sqrt{1-4n^2/N^2} \cos \phi
\hat{H}= (E_{c}/2) \ \hat{n}^2-J(\hat{a}^{\dag}_{l}\hat{a}_{r}+\hat{a}^{\dag}_{r}\hat{a}_{l})
\label{eq:hamiltonianSQ}
\end{equation}
where $\hat{n} = (\hat{a}_{l}^{\dag}\hat{a}_{l}^{\dag}-\hat{a}_{r}^{\dag}\hat{a}_{r})/2$ is half the number difference between the left- and right-hand sites. The operators obey bosonic commutation relations $[\hat{a}_{i},\hat{a}_{j}^{\dag}]=\delta_{ij}$, where $\{i,j\}=\{l,r\}$. The charging energy $E_{c}$ derives from interatomic interactions and $J$  is the hopping energy. In the MF description, which is provided by the Gross-Pitaevskii theory, the Hamiltonian becomes \cite{smerzi97}
\begin{equation}
H= (E_{c}/2) \ n^2-E_{J} \sqrt{1-4n^2/N^2} \cos \phi
\label{eq:hamiltonian}
\end{equation}
where  $\phi= \phi_{r}-\phi_{l}$ is the phase difference between the sites. In MF theory $n$ and $\phi$ take on continuous values (like the amplitude and phase of the electric field in Maxwell's theory of light). If they are promoted to operators obeying $[\hat{\phi},\hat{n}]=i$ then Eq.\ (\ref{eq:hamiltonian}) is equivalent to Eq.\ (\ref{eq:hamiltonianSQ}) if we put $E_{J}=NJ$. In fact, Eq.\ (\ref{eq:hamiltonian}) is more general than the BH model and can also describe the Thomas-Fermi regime \cite{leggett01,P+S,giovanazzi08}. %For simplicity, in what follows we use $n$ interchangeably to represent both the operator and its eigenvalues. 

 The ratio $\lambda = 2E_{J}/E_{c}$  defines three regimes \cite{leggett01}: Fock $\lambda \ll 1$, Josephson $1 \ll \lambda \ll N^2$, and Rabi $\lambda \gg N^2$.
In the Josephson and Fock regimes ($\lambda \ll N^2 $), and when $n \ll N$  (at low energies), $\sqrt{1-4n^2/N^2} \approx 1$ and Eq.\  (\ref{eq:hamiltonian}) becomes the rigid pendulum Hamiltonian. This is the situation we shall assume. The phase space of a pendulum is divided  into two regions by a separatrix: inside the pendulum oscillates back and forth, and outside the pendulum makes complete rotations. 

In the pendulum analogy the first term in the Hamiltonian is kinetic energy, and therefore $E_{c} \propto \hbar^2$. Thus, $\lambda \propto \hbar^{-2}$ \cite{odell01,krahn09}. Small values of $\lambda$ correspond to the quantum regime where there are few energy states below the separatrix, and large values to the semiclassical regime where there are many. In the BH model  $\lambda \propto N$ (there are other possibilities beyond the BH model \cite{wu06,radzihovsky10}), and so $1/\sqrt{N}$ plays the role of Planck's constant. We are interested in how singularities develop as $N$ becomes large, i.e.\ the semiclassical regime \cite{chuchem10,graefe07,shchesnovich08,nissen10}.

Prior to the quench we assume that the tunneling barrier is high,  and the initial state can be approximated by the Fock state $n=0$. However, we can again invoke structural stability to assert that our results will remain qualitatively correct even if the initial state contains a small distribution of Fock states.  At $t=0$ the tunneling barrier is suddenly lowered so that the BJJ goes from the Fock to Josephson regime. 

\begin{figure}[t]
\centering
%\subfigure[]{\includegraphics[width=0.49\columnwidth]{eig.pdf}}
%\subfigure[]{\includegraphics[width=0.49\columnwidth]{DOS.pdf}}
{\includegraphics[width=1.0\columnwidth]{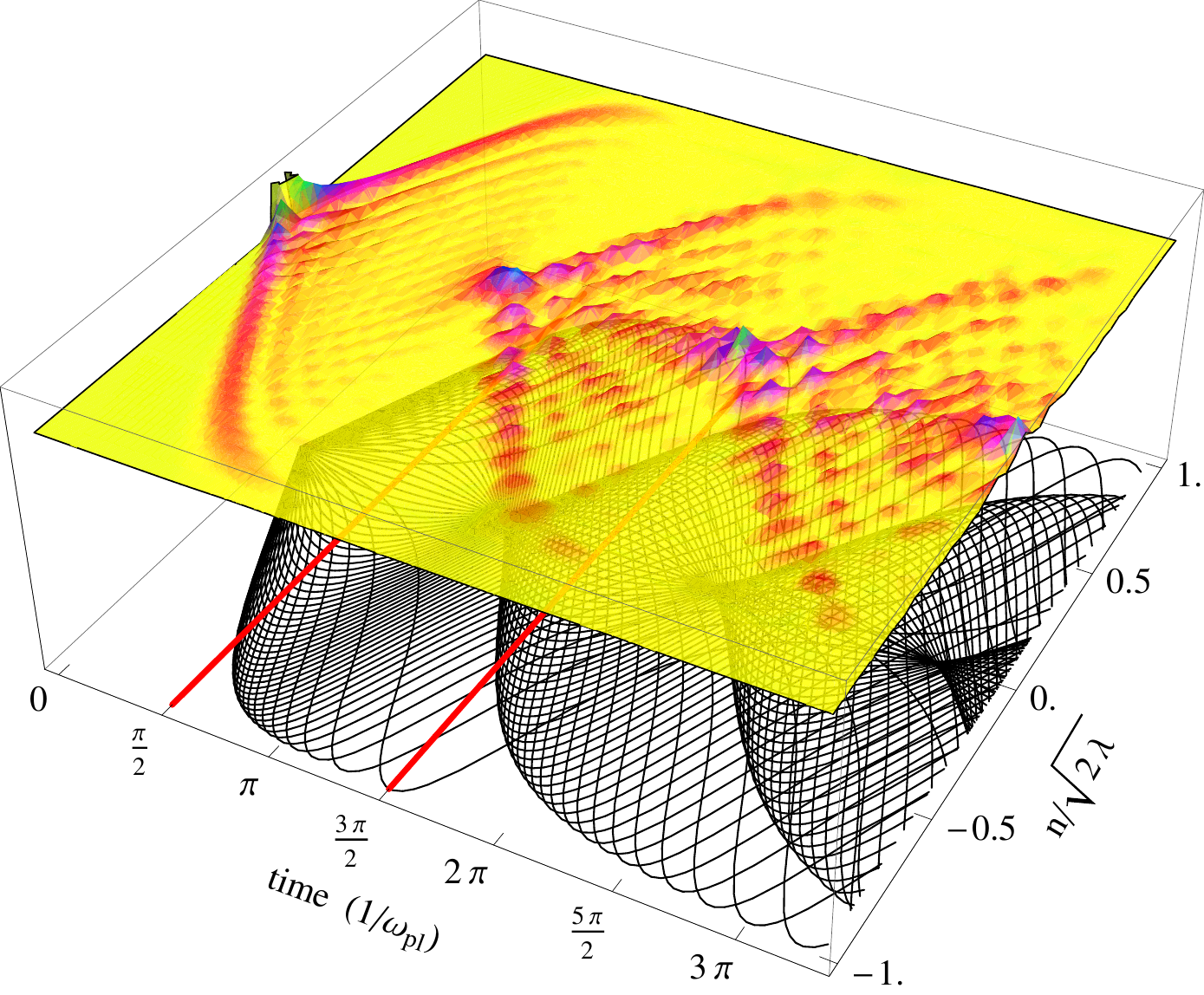}}
\caption{Quantum vs mean-field dynamics following the quench. Time is in units of the inverse plasma frequency, and $n$ is the Fock space coordinate. Above: QMB probability distribution (for $\lambda=200$). Below: MF rays, where each ray has a different initial phase difference $\phi_{0}$, lying in the range $-\pi \ldots \pi$ in steps of $\pi/45$. The two straight red lines indicate the two time slices given in Fig.\ \ref{fig:airyfunc}. }\label{fig:classandquant}
\end{figure}

Consider the MF dynamics. Each MF solution has a perfectly defined phase, but the initial QMB state corresponds to a superposition of all possible phases, and is highly non-classical. To mimic this we form the sum of a large number of MF ``rays'', a sample of which are shown in the lower part of Fig.\ \ref{fig:classandquant}, with initial phases $\phi_{0}$ uniformly distributed over $[0,2\pi)$.  The rays obey Hamilton's (Josephson's) equations $\hbar\dot{\phi}=\partial_{n}H$ and $\hbar\dot{n}=-\partial_{\phi}H$. The solution for the number difference is  \cite{berry99}
\begin{equation}
n(t)=\sqrt{2 \lambda} \sin [\tfrac{1}{2}\phi_{0}] \mathrm{cn} \{\omega_{\mathrm{pl}} t + \mathrm{K} (\sin^{2}[\tfrac{1}{2}\phi_{0}]) \vert \sin^{2} [\tfrac{1}{2} \phi_{0}]\}
\label{eq:ray}
\end{equation}
where $\omega_{\mathrm{pl}}= \sqrt{E_{c}E_{J}}/\hbar$ is the plasma frequency which gives the period of low-lying excitations, and $\mathrm{cn}\{u \vert v \}$ and $\mathrm{K}(v)$ are a Jacobi elliptic function and a complete elliptic integral of the first kind, respectively  \cite{A+Snew}. The characteristic ray pattern in Fig.\ \ref{fig:classandquant} was first calculated in the context of the diffraction of light by ultrasound \cite{lucas32,nomoto51,berry66a}: the cosine potential is not a perfect lens due to its anharmonicity, and rather than focal points gives rise to extended caustics (envelopes of ray families) that proliferate with time. Each new caustic is born as a cusp point (three rays focused to the same point) at the times $t_{m}=m \pi/ \omega_{\mathrm{pl}} $, where $m=0,1,2,\ldots$, and then spreads into a pair of rainbows  (fold caustics due to the focusing of two rays) that gradually move out to $n=\pm\sqrt{2 \lambda}$.  

% commented out so the program runs quicker!

By virtue of Fermat's principle, classical (MF) dynamics can be regarded as a gradient map. The singularities (caustics) of gradient maps are classified by catastrophe theory in terms of their co-dimension $\kappa=\mbox{dimension of space}-\mbox{dimension of caustic}$. According to catastrophe theory, the only structurally stable caustics in the two dimensions formed by combining one-dimensional Fock space with time are cusp points ($\kappa=2$) and fold lines ($\kappa=1$). Catastrophe theory therefore predicts that cusp and fold (rainbow) catastrophes are generic in the time dependent BJJ wave function, and these are exactly what we find in Fig.\ \ref{fig:classandquant}.

\begin{figure}[t]
{\includegraphics[width=1.0\columnwidth]{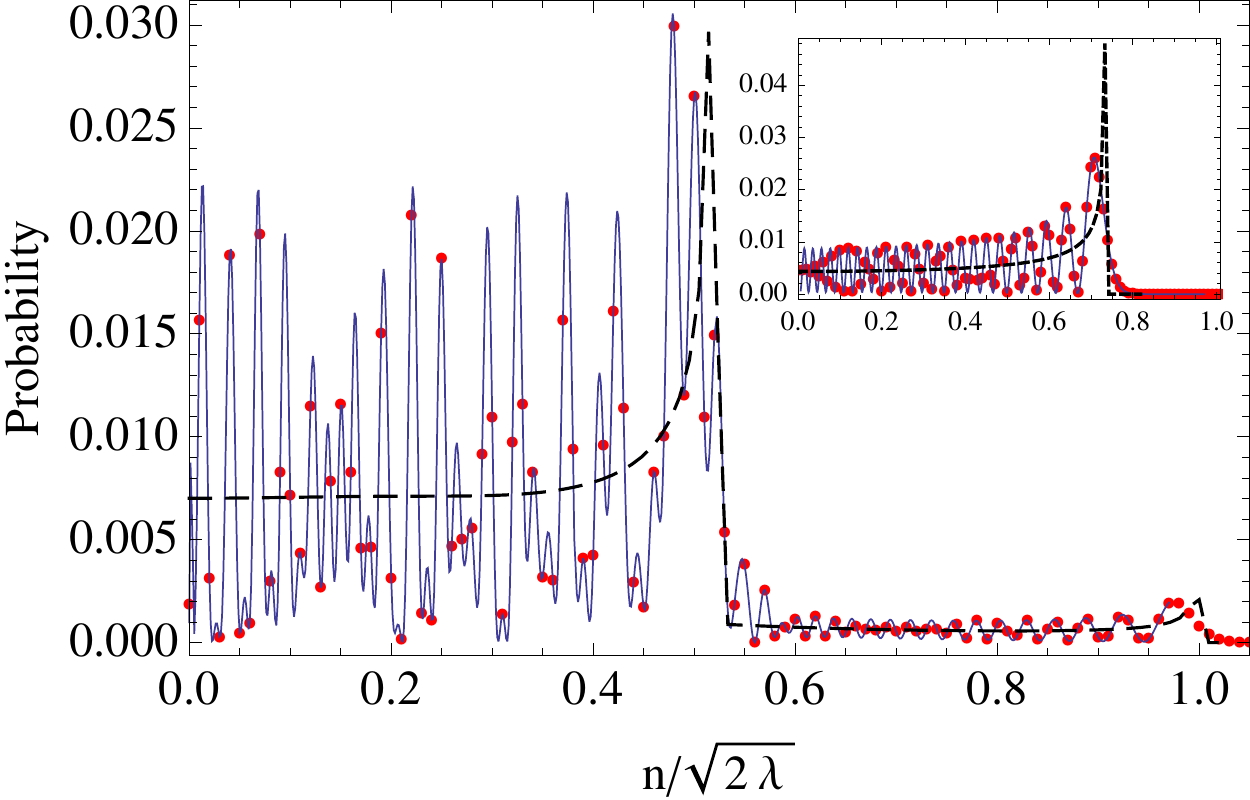}}
\caption{Rainbows in Fock space: number difference probability distribution at the time $t=3 \pi/2 \, \omega_{\mathrm{pl}}^{-1}$ (inset: $t= \pi /2 \, \omega_{\mathrm{pl}}^{-1}$).  Due to the $\pm n$ symmetry of the solutions, only the $+$ve halves are shown. Solid blue curve: semiclassical solution [Eqns (\ref{eq:single}) and (\ref{eq:uniform-wavefunction})] with $\lambda=5000$; red dots: exact QMB solution given by numerical diagonalization;  dashed black curve: MF caustics given by summing 6 million rays [Eq.\ (\ref{eq:ray})].}\label{fig:airyfunc}
\end{figure}

Turning to quantum dynamics, the QMB state takes the form $\vert \Psi(t) \rangle = \sum_{n} C_{n}(t) \vert n \rangle$. The probabilities $\vert C_{n}(t) \vert^2$ for the number difference Fock states $\vert n \rangle$ are plotted in the upper part of Fig.\ \ref{fig:classandquant} and also in Fig.\ \ref{fig:airyfunc}, which displays two time slices. Fig.\ \ref{fig:airyfunc} shows that in the QMB theory the singular MF caustics are replaced by well behaved Airy functions. In order to calculate $\vert \Psi(t) \rangle$
we write it  as a sum over energy eigenstates 
 $\vert \Psi(t) \rangle = \sum_{j} \alpha_{j} \vert E_{j} \rangle \exp[-i E_{j} t/ \hbar ]$. If we express the energy eigenstates as sums over Fock states $\vert E_{j}\rangle = \sum_{n} A_{j,n} \vert n \rangle $, then, by the sudden approximation, the coefficients in the eigensum following the quench are $\alpha_{j}=A_{j,0}$ because the initial state is $\vert n=0 \rangle$. In the semiclassical regime, eigenstates above the separatrix can be dropped from the sum because these states have only an exponentially small amplitude at $n=0$, meaning that $\alpha_{j} \approx 0$.

The characteristic Airy function structure of the rainbows is not at all obvious from the eigensum representation for $\vert \Psi (t)\rangle$. In order to bring it out explicitly we  apply Poisson resummation \cite{berry66} $\sum_{j=0}^{\infty}f(j)=\sum_{m=-\infty}^{\infty} \int_{0}^{\infty} f(j) \exp[2 \pi i mj] dj$, where $f(j)$ refers to the $j^{\mathrm{th}}$ term in the energy eigensum. This transformation is exact and converts the sum over eigenstates into a path integral representation which sums over families of rays \cite{pekeris50}. Each family, labelled by the Maslov index $m$, has geometric significance and corresponds to a rainbow pair that dress the corresponding MF caustics. 
The number  of states $\mathcal{N}$ we need to include in the eigensum is the number below the separatrix,  and in the semiclassical regime $\mathcal{N} \approx \sqrt{2 \lambda}$ \cite{odell01}. By contrast, only two terms are needed in the Poisson sum for the time slice shown in Fig.\ \ref{fig:airyfunc} (and only one term for the inset.)

We now sketch the Poisson resummation of the energy eigensum. Firstly, we replace the Mathieu functions $\vert E_{j} \rangle $ by their WKB approximations \cite{odell01}. We then apply the Poisson summation formula to obtain $\Psi=\langle n \vert \Psi \rangle$ as
\begin{equation}
\Psi= \sum_{m=-\infty}^{\infty} \int_{-1}^{1} \frac{\mathrm{e}^{i \sqrt{\lambda}\mathcal{A}}+\mathrm{e}^{i \sqrt{\lambda}\mathcal{B}}+\mathrm{e}^{i \sqrt{\lambda}\mathcal{C}}+\mathrm{e}^{i \sqrt{\lambda}\mathcal{D}}}{2 \pi D(y,\beta)}d \beta
\label{eq:poisson}
\end{equation}
where the denominator $D = (1-(y^2-\beta)^2)^{1/4}(1-\beta^{2})^{1/4}$, and we have introduced $y = n/\sqrt{\lambda}$ and $\beta = E/\lambda$, which are classical versions of the number difference  and energy  eigenvalues. The energies below the separatrix lie in the range $-1 \leq \beta \leq 1$. The four phases are given by 
\begin{eqnarray}
\mathcal{A} & = & P_{1}-P_{2}+(4m-1)P_{3}-\mbox{$\frac{\beta \omega_{\mathrm{pl}}t}{\sqrt{2}}-\frac{(m-\frac{1}{2})\pi}{\sqrt{\lambda}}$} \\
\mathcal{B} & = & P_{1}-P_{2}+(4m+1)P_{3}-\mbox{$\frac{\beta \omega_{\mathrm{pl}}t}{\sqrt{2}}-\frac{m\pi}{\sqrt{\lambda}}$}\\
\mathcal{C} & = & P_{2}-P_{1}+(4m-1)P_{3}-\mbox{$\frac{\beta \omega_{\mathrm{pl}}t}{\sqrt{2}}-\frac{m\pi}{\sqrt{\lambda}}$} \\
\mathcal{D} & = & P_{2}-P_{1}+(4m+1)P_{3}-\mbox{$\frac{\beta \omega_{\mathrm{pl}}t}{\sqrt{2}}-\frac{(m+\frac{1}{2})\pi}{\sqrt{\lambda}}$}
\end{eqnarray}
where $P_{1}(y,\beta)=y \arccos [y^2-\beta]$, $P_{2}(y,\beta)=2\sqrt{1+\beta} \, \mathrm{E}\{\arccos[y^2-\beta]/2 \, \vert \, 2/(1+\beta)\}$, and $P_{3}(\beta)=P_{2}(0,\beta)$. In these expressions $\mathrm{E} \{ u\vert v\}$ is the Jacobi elliptic integral of the second kind \cite{A+Snew}.

The integrals in Eq.\ (\ref{eq:poisson}) can be evaluated by the uniform approximation \cite{berry66}. When there is only a single stationary point $\beta_{1}$, the phase $\nu$, which represents either $\mathcal{A}$, $\mathcal{B}$, $\mathcal{C}$, or $\mathcal{D}$, can be mapped onto a quadratic function which gives a Gaussian integral and the result
\begin{equation}
\Psi_{\mathrm{single}}=\frac{1}{\sqrt{2 \pi \sqrt{\lambda}\vert \nu''(\beta_{1})\vert}}\frac{\exp[i(\sqrt{\lambda}\nu(\beta_{1})\pm \pi/4)]}{D(y,\beta_{1})} 
\label{eq:single}
\end{equation}
where the $\pm$ sign is fixed by the sign of $\nu''(\beta_{1})$, which is the second derivative of the phase with respect to $\beta$ evaluated at the stationary point. The second derivative of phase $\mathcal{A}$ is $\mathcal{A}'' =  +Q_{1}-Q_{2}+Q_{3}+(1-4m)Q_{4}$, and in terms of this template the others are $\mathcal{B}'' = + -+-+$, $\mathcal{C}''  = -+ -+-$, $\mathcal{D}'' = -+--+$.
%\begin{eqnarray}
%\mathcal{A}'' & = & G_{1}-G_{2}+G_{3}+(1-4m)G_{4} \\
%\mathcal{B}'' & = & G_{1}-G_{2}+G_{3}-(1+4m)G_{4} \\
%\mathcal{C}'' & = & G_{2}-G_{1}-G_{3}+(1-4m)G_{4} \\
%\mathcal{D}'' & = & G_{2}-G_{1}-G_{3}-(1+4m)G_{4} 
%\end{eqnarray}
In these expressions $Q_{1}=\mathrm{F}\{ g \vert h\}/\{4\sqrt{2}h\}$, $Q_{2}=\mathrm{E}\{ g \vert h\}/\{\sqrt{2}(1-\beta^{2})\}$, $Q_{3}=y(2 \beta-y^2)/\{2(1-\beta^2)\sqrt{1-(y^2-\beta)^2}\}$, and $Q_{4}=[ \mathrm{K}(h)-2 \mathrm{E}\{\pi/2 \vert h\}/(1-\beta) ]/\{ 4 \sqrt{2}h \}$, where $\mathrm{F}\{u \vert v\}$  is the Jacobi elliptic integral of the first kind \cite{A+Snew}, and we have denoted $g= \arcsin [\sqrt{(1+\beta-y^2)/(1+\beta)}]$ and $h=(1+\beta)/2$.

On the bright side of a rainbow there are two stationary points $\beta_{1}$ and $\beta_{2}$ giving rise to two-wave interference fringes.
They coalesce and annihilate on the fold line, becoming complex on the dark side where $\Psi_{\mathrm{rainbow}}$ decays exponentially. Near a cusp there are three stationary points which coalesce at the cusp point. For simplicity, we shall concentrate on rainbows.

When there are two stationary points the phase can be mapped onto a cubic polynomial (the simplest of the generating functions prescribed by catastrophe theory \cite{berry81}) and the integral gives an Airy function $\mathrm{Ai}(C)=\int_{-\infty}^{\infty}\exp[i(s^{3}/3+Cs)] ds$. Thus, the uniform approximation for $\Psi$ due to a rainbow is  
\begin{equation}
\begin{split}
& \Psi_{\mathrm{rainbow}}=  \frac{\mathrm{e}^{i \sqrt{\lambda} \bar{\nu}}}{\sqrt{2}} \Bigg\{ \quad \bigg(\frac{1}{D(y,\beta_{1})\sqrt{\nu''(y,\beta_{1})}} \\ & + \frac{1}{D(y,\beta_{2}) \sqrt{-\nu''(y,\beta_{2})}} \bigg) \left( \frac{3 \Delta  \nu }{4 \lambda} \right)^{\frac{1}{6}} \mathrm{Ai} \left[- \left( \tfrac{3}{4} \sqrt{\lambda} \Delta \nu \right)^{\frac{2}{3}} \right] 
\\ & - \mathrm{i}  \bigg(\frac{1}{D(y,\beta_{1})\sqrt{\nu''(y,\beta_{1})}}  -  \frac{1}{D(y,\beta_{2}) \sqrt{-\nu''(y,\beta_{2})}} \bigg) \\ & \times \left( \frac{4}{3 \lambda^{2} \Delta  \nu} \right)^{\frac{1}{6}} \mathrm{Ai}' \left[- \left( \tfrac{3}{4} \sqrt{\lambda} \Delta \nu \right)^{\frac{2}{3}} \right] \Bigg\}  \label{eq:uniform-wavefunction}
\end{split}
\end{equation}
where $\bar{\nu}=[\nu(\beta_{1})+\nu(\beta_{2})]/2$ and $\Delta \nu=\nu(\beta_{2})-\nu(\beta_{1})$. The appearance of the derivative term $\mathrm{Ai}'(C)=d\mathrm{Ai}/dC$ in Eq.\ (\ref{eq:uniform-wavefunction}) is due to a series expansion of the amplitude of Eq.\ (\ref{eq:poisson})  \cite{berry66}. On the fold line ($C=0$) the Airy term dominates the derivative term, but the latter makes a significant contribution away from the caustic.
Note that the $\nu''$ factors in the denominators precisely cancel divergences that would otherwise occur due to the use of WKB approximations for the Mathieu functions.

As seen in Fig.\ \ref{fig:airyfunc}, for $\lambda=5,000$ the semiclassical approximation is indistinguishable from the exact wave function (obtained by numerical diagonalization) which samples it at discrete values of $n$. The only real stationary points at time $t=(\pi /2) \omega_{\mathrm{pl}}^{-1}$ are in the $m=0$ family: the Poisson sum reduces to one term, giving rise to the single rainbow shown in the inset. At  $t=(3 \pi /2) \omega_{\mathrm{pl}}^{-1}$ both the $m=0$ and $m=1$ families contribute: the $m=0$ rainbow has moved to the outside, with the younger $m=1$ rainbow inside it  (the MF singularities occur at $n_{c}/\sqrt{2 \lambda} \approx 0.52$ and $n_{c}/\sqrt{2 \lambda} \approx 0.97$.) The inner rainbow is not a clean Airy function because of interference between the $m=0$ and $m=1$ terms. However, for large  $\lambda$ the  Airy function near the caustic dominates: from Eqns (\ref{eq:single}) and (\ref{eq:uniform-wavefunction}) we find that, \emph{relative} to its background, $\vert \Psi \vert^2$ evaluated on a fold grows as $\lambda^{1/6}$, see Table \ref{tab:lambdadependence}. At cusps the relative height is $\lambda^{1/4}$ \cite{berry81}. Assuming the BH model, this gives peaks in Fock space growing as $N^{1/6}$ for folds and $N^{1/4}$ for cusps, and shows that the macroscopic limit is non-analytic:  in different regions of Fock space $\Psi$ has a different dependence on $N$.

\begin{table}[htbp]
\begin{center}
\begin{tabular}{|r|c|c|c|} \hline
\qquad & $n \ll n_{c}$  (bright) & $n = n_{c}$  & $n \gg n_{c} $ (dark)  \\ \hline
 $\vert \Psi \vert^{2}$ & $\mathcal{O}(\lambda^{-\frac{1}{2}}) \times \cos [ \mathcal{O}(\lambda^{\frac{1}{2}}) ] $  & $\mathcal{O}(\lambda^{-\frac{1}{3}})$ & $\mathcal{O}(\lambda^{-\frac{1}{2}}) \exp [- \mathcal{O}(\lambda^{\frac{1}{2}}) ]$ \\ \hline
\end{tabular}
\end{center}
\caption{Dependence of the number difference probability $\vert \Psi(n) \vert^{2}$ on $\lambda$ as $n$ is varied about a fold caustic (rainbow).}
\label{tab:lambdadependence}
\end{table}

A question of great interest is whether the dynamics of isolated QMB systems can lead to thermodynamic equilibrium  \cite{rigol08,rigol12}.  In Fig.\ \ref{fig:ergodic} we compare the number difference probability distribution computed by averaging the MF solutions over their energy contours in phase space, i.e.\ the microcanonical ensemble (solid red curve), with the result given by averaging the long-time QMB solution over one Josephson period (blue dots). The former can be expressed as \cite{berry99}: 
\begin{equation}
P(n)= \int_{n^2}^{1}\frac{dm}{2 \pi K(m)\sqrt{m(1-m)(m-n^2)(1+n^2-m)}}.
\label{eq:ergodicprobability}
\end{equation}
Their close match indicates that the QMB dynamics is ergodic despite this being an integrable system. This is because with only one degree of freedom the phase space trajectories coincide with energy contours. For comparison, we have also included the distribution  due to a thermal density matrix with Boltzmann weighting (dashed black curve) as might be expected from coupling to a bath (we have assumed a temperature corresponding to the same energy as the coherent state.) The ergodic and thermal distributions are clearly different. The shape of the microcanonical curve in Fig.\ \ref{fig:ergodic} is inherited from the catastrophes: the cusp at $n=0$ is due to the cusps, and the sharp edges are due to rainbows piling up at $n=\pm \sqrt{2 \lambda}$.

The two-mode model [Eqns.\ (\ref{eq:hamiltonianSQ}) and (\ref{eq:hamiltonian})] has proved successful in describing a range of BJJ experiments \cite{albiez05,levy07,gati06,esteve08,gross10,zibold10}. Typical values of $\lambda$ correspond to the semiclassical regime, e.g.\ $4 \times 10^{4}$ \cite{albiez05} and $2 \times 10^{7}$ \cite{levy07}.  Temperatures as low as 20 nK  have been achieved \cite{esteve08}, and so thermal excitation of quasiparticles into higher modes, which can lead to damping \cite{martinez09,zapata03}, can be negligible.  Fock space rainbows (Airy functions) could be seen by measuring $n$ following a quench. The first rainbow pair is generated during the first half plasma period $\pi/\omega_{\mathrm{pl}}$,  and so the system need only propagate for short times: in \cite{albiez05} $\pi/\omega_{\mathrm{pl}}=20$ ms and in \cite{levy07}  $\pi/\omega_{\mathrm{pl}}=6$ ms (in \cite{levy07} undamped ac Josephson oscillations were observed over $150$ ms.) The experiment should be re-run many times for a single propagation time in order to map out an Airy function probability distribution like in Fig.\ \ref{fig:airyfunc}, or for multiple propagation times to map out the probability distribution in Fig.\ \ref{fig:classandquant}.  However,  single atom resolution is not needed to see rainbows because their fringes span many Fock states (the fringes shrink as $\lambda^{-1/3}$ but the number of Fock states excited by the quench grows as $\sqrt{\lambda}$.) Finally, we observe that the microcanonical probability distribution (solid red curve in Fig.\ \ref{fig:ergodic}) is classical and does not require coherence between MF states. 

\emph{Conclusions} Rainbows occur in Fock space when the path integral representation of the many body wave function 
has two coalescing stationary (classical) points. They are places where the MF theory is singular. Second quantization removes the singularities and replaces them with Airy functions which are finite for any finite $N$. Catastrophe theory predicts that such rainbows are both generic and structurally stable: we find they proliferate after a quench in a BJJ. The long-time dynamics following the quench are ergodic and support the hypothesis \cite{rigol12} that even integrable isolated systems can come to microcanonical equilibrium.

\textit{Acknowledgements} Discussions with M.V. Berry, J.H. Hannay and E. Taylor are gratefully acknowledged. This work was funded by NSERC.

\end{document}